\documentclass{appolb}
\usepackage{graphicx}

\begin{document}
\title{Unprecedented studies of the low-energy negatively charged kaons interactions in nuclear matter by AMADEUS 
\thanks{Acta Physica Polonica B}
}
\author{~
\address{~}
\\
{C.~Curceanu$^{1}$, K.~Piscicchia$^{1,2}$,
}
M. Bazzi$^{1}$, C. Berucci$^{3}$, D.~Bosnar$^{4}$, A.M.~Bragadireanu$^{5}$, A.~Clozza$^{1}$, M.~Cargnelli$^{3}$, A. D'uffizi$^{1}$, L.~Fabbietti$^{6}$, C.~Fiorini$^{7}$, F.~Ghio$^{8}$, C.~Guaraldo$^{1}$, M. Iliescu$^{1}$, P. Levi Sandri$^{1}$, J. Marton$^{3}$, D. Pietreanu$^{5}$,  M. Poli Lener$^{1}$, R.~Quaglia$^{7}$,  A. Romero Vidal$^{9}$, E. Sbardella$^{1}$, A. Scordo$^{1}$, H. Shi$^{1}$, D. Sirghi$^{1,5}$, F. Sirghi$^{1,5}$, M. Skurzok$^{10}$, I. Tucakovic$^{1}$,  O. Vazquez Doce$^{6}$, E. Widmann$^{3}$, J.~Zmeskal$^{3}$
\address{$^{1}$INFN Laboratori Nazionali di Frascati, Frascati (Roma), Italy\\
$^{2}$Museo Storico della Fisica e Centro Studi e Ricerche "Enrico Fermi", Roma, Italy\\
$^{3}$Stefan-Meyer-Institut f\"ur subatomare Physik, Vienna, Austria\\
$^{4}$Physics Department, University of Zagreb, Zagreb, Croatia\\
$^{5}$Horia Hulubei National Institute of Physics and Nuclear Engineering (IFIN-HH), Magurele, Romania\\
$^{6}$Excellence Cluster Universe, Technische Universit\"at M\"unchen, Garching, Germany\\
$^{7}$Politecnico di Milano, Dipartimento di Elettronica, Informazione e Bioingegneria and INFN Sezione di Milano, Milano, Italy\\
$^{8}$INFN Sez. di Roma I and Inst. Superiore di Sanita', Roma, Italy\\
$^{9}$Universidade de Santiago de Compostela, Santiago de Compostela, Spain\\
$^{10}$Institute of Physics, Jagiellonian University, Cracow, Poland\\}
}
\maketitle
\begin{abstract}
The AMADEUS experiment aims to provide unique quality data of $K^-$ hadronic interactions in light nuclear targets, in order to solve fundamental open questions in the non-perturbative strangeness QCD sector, like the controversial nature of the $\Lambda(1405)$ state, the yield of hyperon formation below threshold, the yield and shape of multi-nucleon $K^-$ absorption, processes which are intimately connected to the possible existence of exotic antikaon multi-nucleon clusters. AMADEUS takes advantage of the DA$\Phi$NE collider, which provides a unique source of monochromatic low-momentum kaons and exploits the KLOE detector as an active target, in order to obtain  excellent acceptance and resolution data for $K^-$ nuclear capture on H, ${}^4$He, ${}^{9}$Be and ${}^{12}$C, both at-rest and in-flight. 
During the second half of 2012 a successful data taking was performed with a dedicated pure carbon target implemented in the central region of KLOE, providing a high statistic sample of pure at-rest $K^-$ nuclear interactions. For the future dedicated setups involving cryogenic gaseous targets are under preparation.
\end{abstract}

\PACS{13.75.Jz Kaon-baryon interactions,
25.80.Nv Kaon-induced interactions, 21.65.Jk Mesons in nuclear matter}

\section{The AMADEUS scientific case}\label{scie}

The AMADEUS experiment \cite{AMADE,AMADEUS} deals with the study of the low-energy interactions of the negatively charged kaons with light nuclei.
Such type of physics, extremely important for the understanding of the non-perturbative  QCD in the strangeness sector, has important consequences, going from hadron and nuclear physics to astrophysics. In this context, our investigation proceeds along two lines of research, both intimately connected with the antikaon-nucleon potential: the strength of the $K^-$ binding in nuclei and the in-medium modification of the $\Sigma^*$ and $\Lambda^*$ resonances properties.

The investigation of the absorptions of $K^-$ inside the KLOE Drift Chamber (DC)
was originally motivated by the prediction of the formation of deeply bound
kaonic nuclear states \cite{wycech,AkYam}.  Their binding energies and widths
could be determined by studying their decays into baryons and nucleons. In particular,
the dibarionic state $K^- pp$ is expected to decay into $\Lambda p$.

From the experimental point of view, two main
approaches have been used for studying the
$K^- pp$ cluster:
$p$-$p$ and heavy ion collisions
\cite{fopi} \cite{hades1},
and in-flight or stopped $K^-$ interactions
in light nuclei.
For the second, data has been published by the FINUDA \cite{finuda}
and KEK-PS E549 collaborations \cite{kek}.
The interpretation of both results is far from being conclusive,
and it requires an accurate description
of the single and multi-nucleon absorption processes that a
$K^-$ would undergo when interacting with light nuclei.

If exotic antikaon multi-nucleon states exist, then their properties are intimately related to the position and structure of the $\Lambda(1405)$ state. 
The $\Lambda(1405)$ is generally accepted to be a spin $1/2$, isospin $I=0$ and strangeness $S=-1$ negative parity baryon resonance ($J^P=1/2^-$) assigned to the lowest $L=1$ supermultiplet of the three-quark system, together with its spin-orbit partner, the  ($J^P=3/2^-$) $\Lambda(1520)$.
Such state only decays into $(\Sigma \pi)^0$ ($I=0$) through the strong interaction.
Despite the fact that the $\Lambda(1405)$  has been observed in few experiments  and is currently listed as a four-stars resonance in the table of the Particle Data Group (PDG) \cite{PDG}, its nature still remains an open issue.
The three quark picture ($uds$) meets some difficulties to explain both the observed $\Lambda(1405)$ mass and the mass splitting with the $\Lambda(1520)$.
The low mass of the $\Lambda(1405)$ can be explained in a five quark picture, which, however, predicts more unobserved excited baryons.
In the meson-baryon picture the  $\Lambda(1405)$ is viewed as a $\overline{K}N$ quasi-bound $I=0$ state, embedded in the $\Sigma \pi$ continuum, emerging in coupled-channel meson-baryon scattering models \cite{dali3}.
A complete review of the broad theoretical work, and references to the experimental literature can be found in \cite{Hyodo}. The $\Lambda(1405)$ production in $\overline{K}N$ reactions is of particular interest due to the prediction, in chiral unitary models \cite{kaiser,oset,oller}, of two poles emerging in the scattering amplitude (with $S=-1$ and $I=0$) in the neighborhood of the  $\Lambda(1405)$ mass. One pole is located at higher energy with a narrow width and is mainly coupled to the $\overline{K}N$ channel, while a second lower mass and broader pole is dominantly coupled to the $\Sigma \pi$ channel \cite{weise2012}, and both contribute to the final experimental invariant mass distribution \cite{oller,nacher}.
The $\Sigma^0 \pi^0$ decay channel, which is free from the $I=1$ contribution and from the isospin interference term, is expected to contain the cleanest information on the $\Lambda(1405)$ state and represented for such reason the the starting point of such analysis. As the $\Sigma^+ \pi^-$ invariant mass spectrum can be reconstructed with much better resolution, such channel is also under investigation.

The $K^-$ nuclear absorption \emph{at rest} on ${}^4He$ and ${}^{12}C$ was explored, through the $\Sigma^\pm \pi^\mp$ channels, in bubble chamber \cite{riley} and emulsion experiments \cite{rif1,rif2}. In the present study the contribution of the \emph{in flight} $K^-$ nuclear absorption to the  $\Sigma \pi$ invariant mass spectra was evidenced. The capture of low momentum ($\sim 120$ MeV) kaons from DA$\Phi$NE on bound protons in  ${}^4He$ and ${}^{12}C$ allows to access a higher invariant mass region, above the threshold imposed by the proton binding energy, otherwise inaccessible.

Unveiling the shape of the $\Lambda(1405)$ state requires the knowledge of the non-resonant transition amplitude below threshold. 
The non-resonant versus resonant formation rate could be measured, for the first time, for different isospin states, for a broad range of light nuclei. Such study is presently  being performed, for the $I=1$ $K^- \, n$ interaction, through the investigation of the $\Lambda \pi^-$ correlated production.

\section{The DA$\Phi$NE collider and the KLOE detector}

DA$\Phi$NE \cite{dafne} (Double Anular $\Phi$-factory for Nice Experiments) is a double ring $e^+ \, e^-$ collider, designed to work at the center of mass energy of the $\phi$ particle $m_\phi = (1019.456 \pm 0.020) \, \mathrm{MeV/c^2}$.
The $\phi$ meson decay produces charged kaons (with BR($K^+ \, K^-$) = $48.9 \pm 0.5 \%$) with low momentum ($\sim 127$ $\mathrm{MeV/c}$) which is ideal either to stop them, or to explore the products of the low-energy nuclear absorptions of $K^-$s. The back-to-back topology, characterizing the kaons pair production, is extremely useful for the extra\-polation of non identified kaon tracks. 

The KLOE detector  \cite{kloe} is centered around the interaction region of DA$\Phi$NE. KLOE is characterized by a $\sim 4\pi$ geometry and an acceptance of $\sim98\%$; it consists of a large cylindrical Drift Chamber (DC) and a fine sampling
lead-scintillating fibers calorimeter, all immersed in an axial magnetic field of 0.52 $\mathrm{T}$, provided by a superconducting solenoid.
The DC \cite{kloedc} has an inner radius of 0.25 $\mathrm{m}$, an outer radius of 2 $\mathrm{m}$ and a length of 3.3 $\mathrm{m}$. The DC entrance wall composition is 750 $\mathrm{\mu m}$ of carbon fibre and 150 $\mathrm{\mu m}$ of aluminium foil.

Dedicated GEANT MC simulations  of the KLOE apparatus were performed to estimate the percentages of $K^-$ absorptions in the materials of the DC entrance wall (the $K^-$ absorption physics were treated by the GEISHA package). Out of the total fraction of captured kaons, about 81$\%$ results to be absorbed in the carbon fibre component and the residual 19$\%$ in the aluminium foil.
The KLOE DC is filled with a mixture of helium and isobutane (90$\%$ in volume $^4$He and 10$\%$ in volume $C_4H_{10}$). 

The chamber is characterized by excellent position and momentum re\-solutions. 
Tracks are reconstructed with a resolution in the transverse $R-\phi$ plane of
$\sigma_{R\phi}\sim200\,\mathrm{\mu m}$ and a resolution along the z-axis of $\sigma_z\sim2\,\mathrm{mm}$.
The transverse momentum resolution for low momentum tracks ($(50<p<300) \mathrm{MeV/c}$)
is $\frac{\sigma_{p_T}}{p_T}\sim0.4\%$.
The KLOE calorimeter \cite{kloeemc} is composed of a cylindrical barrel and two endcaps,
providing a solid angle coverage of 98\%.
The volume ratio (lead/fibres/glue=42:48:10) is optimized for
a high light yield and a high efficiency for photons in the range
(20-300) MeV/c. The position of the cluster along the fibres can be obtained with a resolution $\sigma_{\parallel} \sim 1.4\, \mathrm{cm}/\sqrt{E(\mathrm{GeV})}$. The resolution in the orthogonal direction is  $\sigma_{\perp} \sim 1.3\, \mathrm{cm}$. The energy and time resolutions for photon clusters are given by $\frac{\sigma_E}{E_\gamma}= \frac{0.057}{\sqrt{E_\gamma (\mathrm{GeV})}}$ and 
$\sigma_t= \frac{57 \, \mathrm{ps}}{\sqrt{E_\gamma (\mathrm{GeV})}} \oplus 100 \,\, \mathrm{ps}$.

As a step 0 of the presented work, we analysed the 2004-2005 KLOE collected data, for which the $dE/dx$ information of the reconstructed tracks is available ($dE/dx$ represents the truncated mean of the ADC collected counts due to the ionization in the DC gas). An important contribution of in-flight $K^-$ nuclear captures, in different nuclear targets from the KLOE materials, was evidenced and characterized, enabling to perform invariant mass spectroscopy of in-flight $K^-$ nuclear captures \cite{bormio}. In order to increase the statistics and as an essential interpretation tool, AMADEUS step 1 consisted in the realization of a dedicated pure carbon target, implemented in the central region of the KLOE detector, providing a high statistics sample of pure at-rest $K^-$ nuclear interaction, as described in the next section.

\section{Data samples}

\begin{figure}[htb]
\centerline{%
\includegraphics[width=11.5cm]{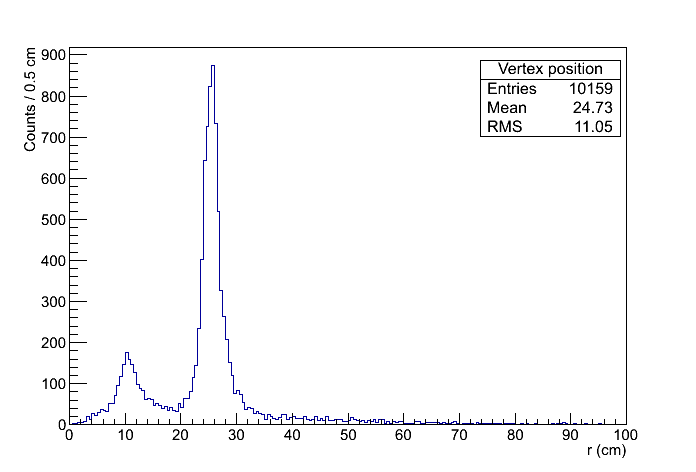}}
\caption{\em Radial position distribution $\rho_\Lambda$, of the $\Lambda$ decay vertex, for 2004-2005 KLOE collected data.}
\label{kloe2004}
\end{figure}

Two different data samples are presently under study. A first sample was collected by KLOE in the period 2004-2005 (2.2 fb$^{-1}$ total statistics). In this case the DA$\Phi$NE beam pipe, and the materials of the KLOE detector are used as active targets. The topology of these data is shown in figure \ref{kloe2004}, representing the radial position ($\rho_\Lambda$) of the $\Lambda(1116)$ decay vertex (see Section \ref{lambda}). Four components are distinguishable, from inside to outside  we recognize $K^-$ absorptions in the DA$\Phi$NE beryllium sphere ($\sim$ 5 cm), the DA$\Phi$NE aluminated beryllium pipe ($\sim$ 10 cm), the KLOE DC entrance wall (aluminated carbon fibre $\sim$ 25 cm) and the long tail originating from $K^-$ interactions in the gas filling the KLOE DC (25-200 cm). Extremely rich experimental information is contained in this sample, with $K^-$ hadronic interactions, both at-rest and in-flight, in a variety of light nuclear targets (H, ${}^4$He, ${}^9$Be and ${}^{12}$C), which, of course, turned into a challenging analysis and interpretation effort.

Motivated by the obtained results, a high purity carbon target (graphite) was realized in summer 2012 and installed inside the KLOE DC, between the beam pipe and the DC entrance wall. The target was realized with the aim to confirm the in-flight component hypothesis in the 2004-2005 data, and to collect a higher statistics for the study of the low-energy $K^- \,  {}^{12}C$ hadronic interaction.

\begin{figure}[htb]
\centerline{%
\includegraphics[width=11.5cm]{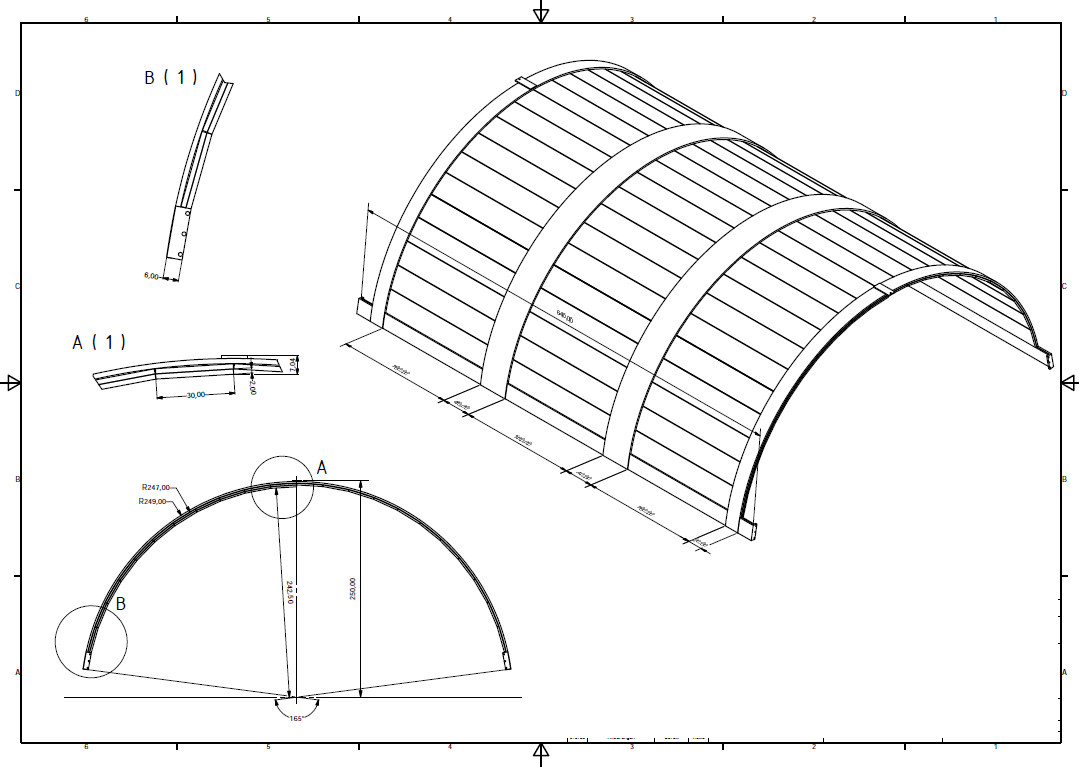}}
\caption{\em Project of the graphite target inserted between the beam pipe and the KLOE DC entrance wall.}
\label{target}
\end{figure}

\begin{figure}[htb]
\centerline{%
\includegraphics[width=11.5cm]{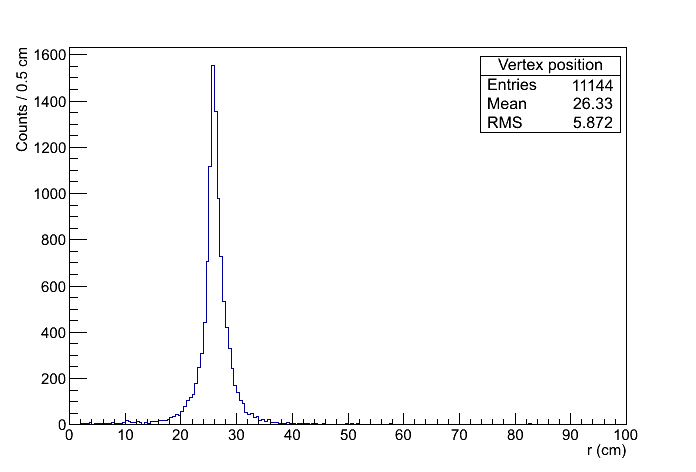}}
\caption{\em Radial position distribution of $\rho_\Lambda$ for carbon target events.}
\label{target1}
\end{figure}

The geometry of the target was optimized by means of a GEANT3 MC simulation.
The final configuration was a three-sector half-cylinder ($\rho=$ 0.95g/cm$^3$) supported by an aluminium frame, with a length of 600 mm and a mean radius of 24,7 mm. The thickness varied from 4 to 6 mm, in order to optimize the percentage of stopped kaons, taking into account the $\phi$ boost, derived by the limited crossing angle of the e$^+$ e$^-$ beams. 
The half-cylindrical configuration was chosen in order to take advantage of the $K^+$ tagging in the opposite direction.  
The final project of the realized target is shown in figure  \ref{target}. 
We took data from 6 November to 14 December 2012, for a total integrated luminosity of $\sim$90 pb$^{-1}$.  Up to now we analysed a sample of 37 pb$^{-1}$ reconstructed data. The topology of the reconstructed $\Lambda$ decay vertices radial position is shown in figure \ref{target1}, as expected, the majority of kaons are stopped in the target.

\section{Preliminary results of the data analyses}\label{prelim}

The investigation of the negatively charged kaons interactions in nuclear matter is performed through the reconstruction of hyperon-pion and hyperon-nucleon/nucleus correlated pairs productions, following the $K^-$ absorptions in H, ${}^4$He, ${}^9$Be and ${}^{12}$C.

The investigation of the $K^-$ multi-nucleon absorptions and the properties of possible antikaon multi-nucleon bound states proceeds through the analysis of the $\Lambda-p$, $\Lambda-d$ (which are the expected decay channels of eventual $K^-pp$ and $K^-ppn$ clusters) and $\Lambda-t$ correlations; this last channel is, in particular, extremely promising for the search and characterization in different nuclear targets of the extremely rare four nucleon absorption process. 

The search for the $\Lambda(1405)$ is performed through its decay in $\Sigma^0\pi^0$ (purely isospin I=0) and $\Sigma^+\pi^-$ (also the analysis of the $\Sigma^-\pi^+$ decay channel started recently with  a characterization of neutron clusters in the KLOE calorimeter). The line shapes of the three combinations $(\Sigma\pi)^0$ were recently obtained, for the first time, in a single (photoproduction) experiment by \cite{moriya}; as the line-shapes of the three invariant mass spectra were found to be different, a comparative study with $K^-N$ production is of extreme interest. Moreover a precise measurement of the $\frac{\Sigma^+\pi^-}{\Sigma^-\pi^+}$ production ratio in different targets can unveil the nature of the $\Lambda^*$ state, by observing modifications of its parameters as a function of the density  \cite{wycech1,ohnishi}.  

To conclude, Given the excellent resolution for the  $\Lambda \pi^-$ invariant mass, the analysis of the $\Lambda \pi^-$ (isospin I=1) production, both from direct formation process and from internal conversion of a primary produced $\Sigma$ hyperon ($\Sigma \, N \rightarrow \Lambda \, N'$) is presently ongoing. Our aim is to measure, for the first time, the module of the non-resonant transition amplitude (compared with the resonant $\Sigma^{*-}$) below threshold. 

Preliminary results from the $\Lambda-p$ and $\Sigma^0\pi^0$ studies will be presented in Sections \ref{lp} and \ref{s0pi0} respectively.

\subsection{The $\Lambda(1116)$ selection}\label{lambda}

The presence of a hyperon always represents the signature of $K^-$ hadronic interaction inside the KLOE materials. Most of the analysis introduced in Section \ref{prelim} then start with the identification of a $\Lambda(1116)$, through the reconstruction of the $\Lambda \rightarrow p + \pi^-$ (BR = 63.9 $\pm 0.5 \%$) decay vertex.

\begin{figure}[ht]
\centering

\begin{tabular}{rl}
\hspace{-1.cm}
\mbox{\includegraphics*[height=6.cm,width=7cm]{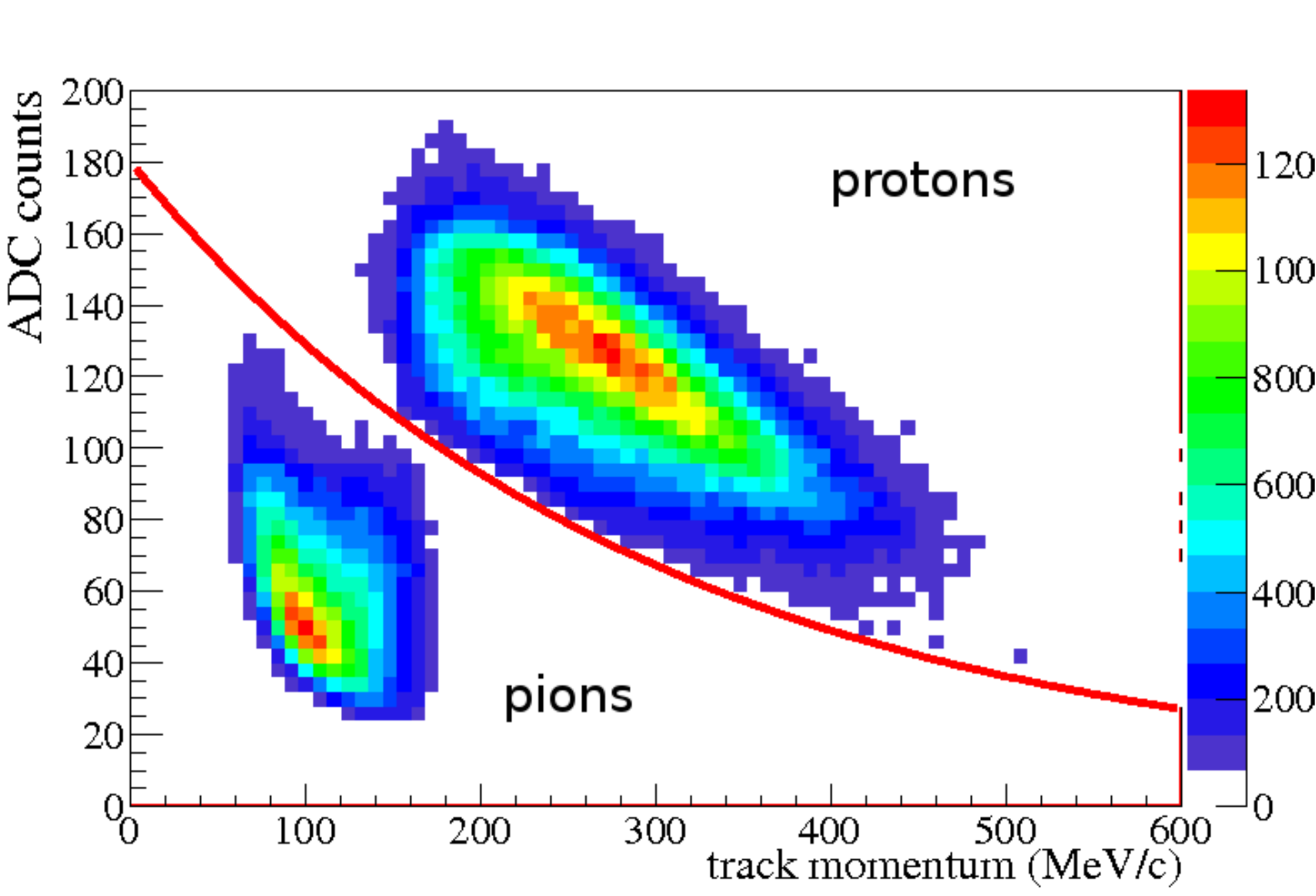}}
\hspace{-.1cm}
\mbox{\includegraphics*[height=6.cm,width=6cm]{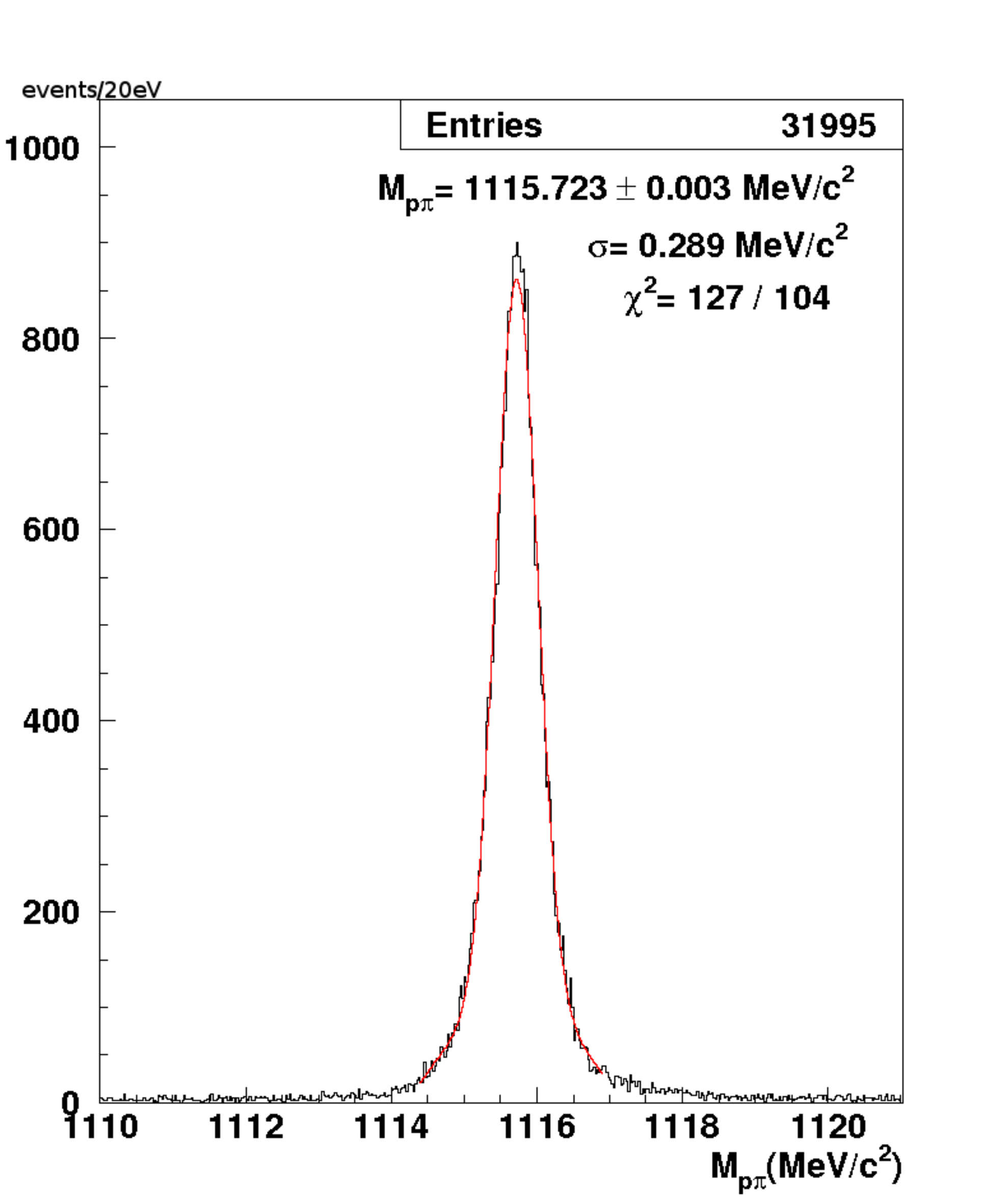}}
\end{tabular}


\caption{\em Left: $dE/dx$ (in ADC counts) vs. momentum for the selected proton (up) and pion (down) tracks in the final selection. The proton selection functions is displayed in red. Right: $m_{p\pi^-}$ invariant mass spectrum for the final selected pion-proton pairs.}
\label{Lmass}       
\end{figure}

In order to reduce the copious background from the three-body $K^{\pm}$ decays   ($K^\pm \rightarrow \pi^{\pm} \pi^{\pm} \pi^{\mp}$) we had
to refine the search criteria for the proton.
To this aim a cut on $dE/dx$ was optimized by characterizing
proton tracks with an associated cluster
in the calorimeter.
The signature of a proton in the calorimeter is clean,
since the corresponding signal is well-separated by the signal generated by the pions.
We were tuning the cut on these protons and then
applied it to all protons, i.e. including those which have the last DC measurement near the calorimeter (reaching it) but have no associated cluster. This last requirement enables to include in the selection low momentum protons (lower than $p \sim 250$ MeV/c) not producing an observable signal in the calorimeter. 
In figure \ref{Lmass} left
the $dE/dx$ versus momentum scatterplot for the finally selected protons
is shown, the function used for the selection of protons is displayed in red.
 The typical signature of pions in $dE/dx$ versus momentum can be also seen in figure \ref{Lmass} left illustrating the efficient rejection of 
$\pi^+$ contamination in a broad range of momenta.

A minimum track length of 30 cm is required, and a common vertex is searched for all the $p-\pi^-$ pairs in each event.
When found, the common vertex position is added as an additional constraint for the track refitting. The module of the momentum and the vector cosines are redefined for both tracks,
taking into account for the energy loss in the gas and the various crossed materials (signal and field wires, DC wall, beam pipe) when tracks are extrapolated back through the detector.
In addition protons are required to have a momentum $p$ $>$ 130 MeV/c, in order to remove the contamination from two body kaon decays ($K^+ \rightarrow \pi^+ \pi^0$), when the charge of the $\pi^+$ is misidentified.
As a final step for the identification of $\Lambda$ decays, the vertices are cross checked with quality cuts using the minimum distance between tracks (minimum distance $<$ 3.2 cm) and the chi-square of the vertex fit.

A spatial resolution below 1 mm is achieved for vertices found inside the DC volume (evaluated with Monte Carlo).
The invariant mass $m_{p\pi^-}$, calculated under the $p$ and  $\pi^-$ mass
hypothesis, is represented
in figure \ref{Lmass} right. The Gaussian fit gives a mass of 1115.723 $\pm$ 0.003 MeV/c$^2$ and an excellent sigma of 0.3 MeV/c$^2$, this confirms the unique performances of KLOE for charged particles (the systematics, depending on the momentum calibration of the KLOE setup, are presently under evaluation).


Cuts on the radial position ($\rho_\Lambda$) of the $\Lambda$ decay vertex were optimized in order to separate the two samples of $K^-$ absorption events occurring in the DC wall and the DC gas: $\rho_\Lambda = 25 \pm 1.2 $ cm and $\rho_\Lambda > 33$ cm, respectively. The $\rho_\Lambda$ limits were set based on MC simulations and a study of the $\Lambda$ decay path. In particular, the $\rho_\Lambda = 25 \pm 1.2 $ cm cut guarantees, for the first sample, a \emph{contamination} of $K^-$ interactions in gas as low as $(5.5^{+1.3}_{-1.8} \%)$.

\subsection{$\Lambda$p correlation study}
\label{lp}

\begin{figure}[htb]
\centerline{%
\includegraphics[width=9.cm]{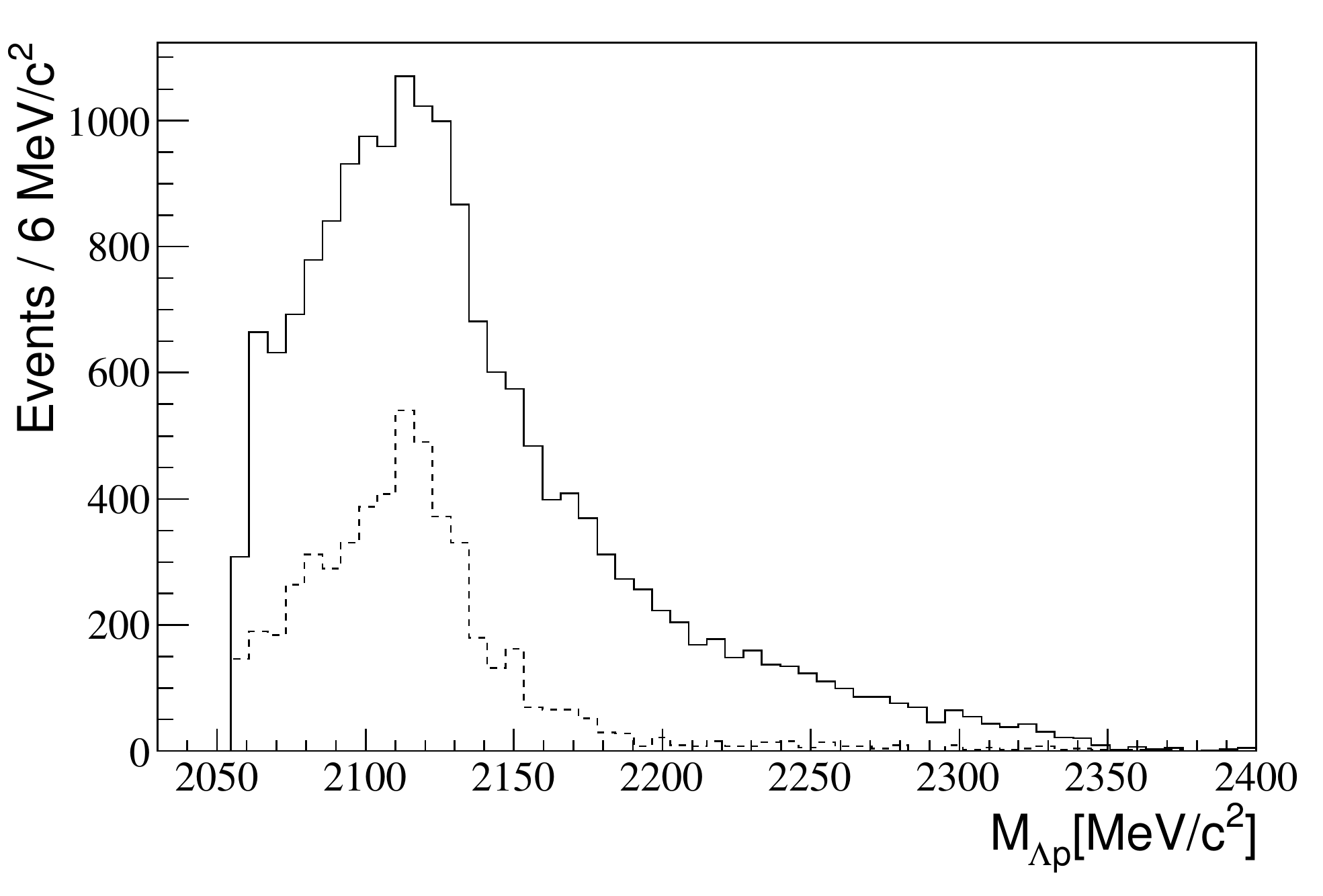}}
\caption{\em $\Lambda$p invariant mass for the inclusive $\Lambda$p selection (continuous line) and for events with an extra $\pi^-$ detected (dashed line, arbitrary normalization).}
\label{plot}
\end{figure}

The reconstruction of $\Lambda-p$ correlated pairs continues with the search for an extra proton correlated with the $\Lambda$ decay vertex. The extra proton search follows the same prescription described in Section \ref{lambda}. The primary interaction vertex is then found extrapolating backwards the $\Lambda$ path and the extra proton track. The resolution in the $\Lambda-p$ invariant mass ($\sigma_{m\Lambda p}=1.10 \pm 0.03$ MeV) is increased by improving the reconstruction of the proton four-vector; moreover, the material where the absorption takes place can be more precisely determined. The invariant mass distribution for the final selected  $\Lambda-p$ inclusive events is shown in  figure \ref{plot} for $K^-$ captures in the KLOE DC volume. The enhancement characterizing the low invariant mass region is expected to be populated mainly by events coming from the single nucleon absorption. In this case the kaon interacts with one nucleon, producing an hyperon-pion pair. The extra detected proton is a fragment of the residual nucleus \cite{velde} (thus initially not participating in the interaction) or comes from the $\Sigma$/$\Lambda$ nuclear conversion process \cite{roosen} taking place in the residual nucleus. Superimposed on the conversion processes, the shoulder at 2130 MeV could be caused by the $\Lambda p \, - \, \Sigma^0 p$ cusp effect, as recently observed by the COSY-ANKE experiment \cite{cosy}. As anti \- cipated in Section \ref{scie} the knowledge of the shape of the $\Lambda-p$  invariant mass spectrum produced by the single and multi-nucleon absorption processes, is of great importance in the search for kaonic bound clusters.

The dominant contribution to the low invariant mass region of figure \ref{plot} comes from the mesonic processes, and can be better characterized. A negatively charged pion correlated with the primary absorption vertex is searched for, following the same procedure used for the identification of the extra proton. This extra pion is found in approximately 1/6 of the total $\Lambda-p$ events. The invariant mass for this subsample is represented by the dashed line in figure \ref{plot}, with an arbitrary normalization, in order to be compared with the full spectrum. With the requirement of the presence of a pion track, the contribution from the high energy tail loses strength, when compared with the inclusive selection.

The interpretation of the KLOE data will allow both qualitative and quantitative characterizations of the $K^-$ single and multi-nucleon absorption processes to be added to the $p$-$p$ and heavy ion collisions informations. From the lambda-nucleon/nucleus channel analyses the fraction of  single versus multi-nucleon absorption processes can be extracted in $^4He$, $^{9}Be$ and $^{12}C$, the rates of nuclear $\Sigma$/$\Lambda$ conversion, for the given energy range, can be obtained as well. such studies will give important inputs to the sub threshold modification of the antikaon-nucleon interaction.

\subsection{The $\Sigma^0 \pi^0$ identification}\label{s0pi0}

The selection of $\Sigma^0 \pi^0$ events proceeds, after the $\Lambda(1116)$ identification, with the search for three additional in-time photon clusters. We will indicate as $\gamma_3$ the photon coming  from the $\Sigma^0$ decay, $\gamma_1$ and $\gamma_2$ will then represent the photons from $\pi^0$ decay, according to the reaction:

\begin{equation}
K^- p \rightarrow \Sigma^0 \pi^0 \rightarrow (\Lambda \gamma_3) \, (\gamma_1 \gamma_2) \rightarrow (p\pi^-) \gamma_1 \gamma_2 \gamma_3
\end{equation}
A pseudo-chisquare minimization is performed, searching for three neutral clusters in the calorimeter ($E_{cl}>20$ MeV), in time from the decay vertex position of the $\Lambda(1116)$ ($\rho_\Lambda$) ($\chi_t^2=(t_i-t_j)^2/\sigma_t^2$ where $t_i$ is the i-\emph{th} cluster time subtracted by the time of flight in the speed of light hypothesis).
According to dedicated MC simulations a cut was optimized on this variable $\chi_t^2 \le 20$.

Once the three candidate photon clusters are chosen, their assignment to the correct triplet of photons, ($\gamma_1,\gamma_2,\gamma_3$), is based on a second pseudo-chisquare minimization  ($\chi_{\pi\Sigma} ^2$). $\chi_{\pi\Sigma} ^2$ involves both the $\pi^0$ and $\Sigma^0$ masses. $\chi^2_{\pi\Sigma}$ is calculated for each possible combination and the minimizing triplet is selected. The $\chi^2_{\pi\Sigma} \le 45$ cut was optimized based on MC simulations.

According to true MC information the algorithm has an efficiency of $(98\pm1) \%$ in recognizing photon clusters and an efficiency of $(78\pm1)\%$ in distinguishing the correct $\gamma_1 \gamma_2$ pair ($\pi^0$ decay) from $\gamma_3$.

A check is performed on the clusters energy and distance to avoid the selection of splitted clusters  (single clusters in the calorimeter erroneously recognized as two clusters) for $\pi^0$s. Cluster splitting is found to not affect significantly the sample.

\begin{figure}[htb]
\centerline{%
\includegraphics[width=9.cm]{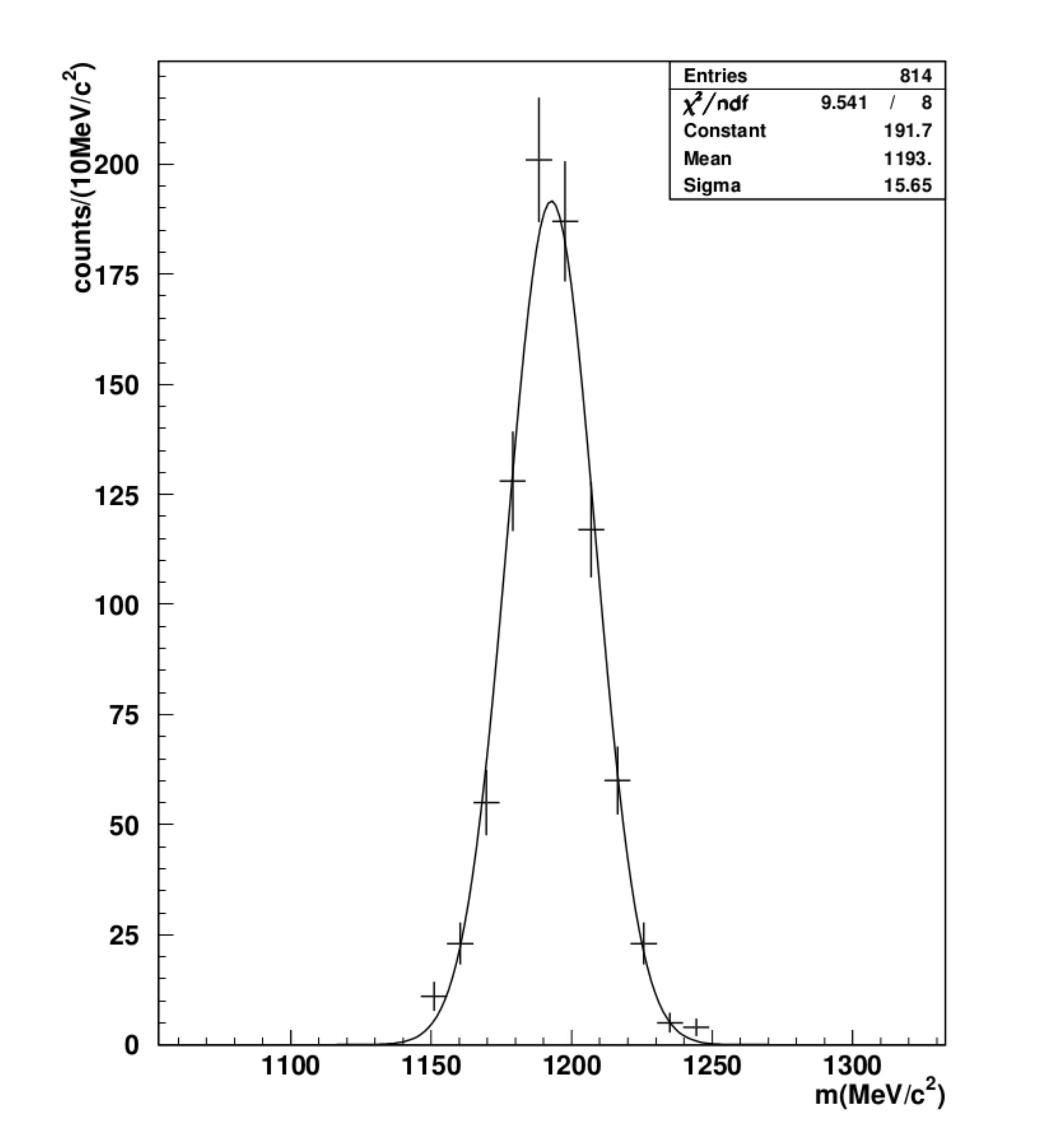}}
\caption{\em $m_{\Lambda\gamma 3}$ invariant mass distribution, together with a Gaussian fit}
\label{s0}
\end{figure}

In figure \ref{s0} is shown the obtained invariant mass $m_{\Lambda\gamma 3}$ (for absorptions in the gas) together with a Gaussian fit. The resolution in the $m_{\Lambda \gamma 3}$ invariant mass is  $\sigma_{m_{\Lambda \gamma 3}} \sim 15$ MeV/c$^2$.
The resolutions on $\rho_\Lambda$ for the final selected $\Lambda$s are $\sigma_{\rho_\Lambda} \sim 0.20$ cm (DC wall) and $\sigma_{\rho_\Lambda} \sim 0.13$ cm (DC gas). The resolutions on the $\Lambda$ momentum are  $\sigma_{p_{\Lambda}} \sim 4.5$ MeV/c (DC wall) and $\sigma_{p_{\Lambda}} \sim 1.9$ MeV/c (DC gas). Each quoted mean resolution corresponds to  Gaussian fits to the distributions of the originally generated true-MC quantities subtracted by the reconstructed ones. The better resolution for the measured variables corresponding to $K^-$ hadronic interactions in the gas filling the KLOE DC is a consequence of the charged particles energy loss, mainly in the material of the DC entrance wall, particularly important for protons.

\section{Conclusions and perspectives}

The broad experimental program of AMADEUS, dealing with the non-perturbative QCD in the strangeness sector, is supported by the quest for high precision and statistics measurements, able to set more stringent constraints on the existing theoretical models. We demonstrated the capabi\-lities of the KLOE detector\footnote{The results presented in this paper are based on a preliminary analysis of members of the AMADEUS Collaboration on the data collected by the KLOE/KLOE2 collaboration with the KLOE detector. The permission to access the data collected with the KLOE detector has been granted by the KLOE/KLOE2 Collaboration to members of the AMADEUS Collaboration with the purpose of studying the feasibility of some measurements in the field of the Physics of kaon-nucleon interactions. At this stage no scientific responsibility for the results presented in this paper can be given to the members of the KLOE/KLOE2 collaboration.} to perform high quality physics (taking advantage of the unique features of the DA$\Phi$NE factory) in the open sector of strangeness nuclear physics. Our investigations, presently spread on a wide spectrum of physical processes, represent one of the most ambitious and systematic efforts in this field.

We are presently ongoing with the data analyses, both for the 2004-2005 KLOE data, and for the the dedicated carbon-target data collected in 2012. In parallel, from the experimental point of view, we are considering the preparation of a dedicated setup, to explore more in detail and with an even higher precision, low-energy kaons interactions with targets going from hydrogen and helium, to lithium and beryllium.

The studies of the low-energy kaon-nuclei interactions performed by AMADEUS are having implications going from nuclear and particle physics to astrophysics (equation of state of neutron stars). Is there any place for strangeness in the stars or any other form of stable matter? AMADEUS will help to answer such questions.

\section*{Acknowledgements}

We thank F. Bossi, S. Miscetti, E. De Lucia, A.
Di Domenico, A. De Santis and V. Patera for the guidance in performing
the data analyses, and
all the KLOE Collaboration and the DA$\Phi$NE staff for the fruitful collaboration. We are grateful to P. Moskal and M. Silarski for very fruitful discussions and suggestions which added as well new fascinating perspectives.  Thanks to Doris Stueckler and Leopold Stohwasser for the technical realization of the carbon-target.

We acknowledge the Croatian Science Foundation under Project No. 1680,
and the Research Grant PRELUDIUM 6 No 2013/11/N/ST2/04152, Poland.

Part of this work was supported
by the European Community-Research Infrastructure Integrating Activity ``Study of Strongly 
Interacting Matter'' (HadronPhysics2, Grant Agreement No. 227431, and HadronPhysics3 (HP3)
Contract No. 283286) under the EU Seventh Framework Programme.

\end{document}